\documentclass[superscriptaddress,
twocolumn,secnumarabic,nofootinbib]{revtex4-1}
\usepackage{amsmath}
\usepackage[utf8]{inputenc}
\usepackage[T1]{fontenc}

\pdfoutput=1

\usepackage{bm}
\usepackage{times}
\usepackage{braket}
\usepackage{color,graphicx}
\usepackage[utf8]{inputenc}
\usepackage[T1]{fontenc}
\usepackage[export]{adjustbox}

\usepackage{enumitem}

\usepackage{verbatim}

\usepackage{float}

\usepackage{colortbl}
\usepackage{longtable}
\usepackage{tabularx}
\usepackage{makecell}

\usepackage{multirow}

\usepackage{xcolor}

\usepackage{lipsum}

\usepackage[normalem]{ulem}

\definecolor{LightGray}{gray}{0.96}
\definecolor{Gray}{gray}{0.94}
\definecolor{nicered}{rgb}{0.7,0.1,0.1}
\definecolor{nicegreen}{rgb}{0.1,0.5,0.1}

\usepackage[colorlinks,
            linkcolor=black,
            filecolor=black,
            anchorcolor=black,
            urlcolor=black,
            citecolor=blue,
            bookmarks=false,
            ]{hyperref}

\allowdisplaybreaks

\setlongtables

\begin{document}

\title{\boldmath Are the new particles heavy or light in $b \to s E_{\mathrm{miss}}$?}

\author{Quan-Yi Hu} 
\email{huquanyi@gxnu.edu.cn}
\affiliation{Department of Physics, Guangxi Normal University, Guilin 541004, Guangxi, China}
\affiliation{Guangxi Key Laboratory of Nuclear Physics and Technology, Guangxi Normal University, Guilin 541004, Guangxi, China}
\affiliation{School of Physics and Electrical Engineering, Anyang Normal University, Anyang 455000, Henan, China}
  
\begin{abstract}
In this work, we study the $B^+\to K^+ E_{\mathrm{miss}}$, $B^0\to K^{*0} E_{\mathrm{miss}}$, and $\Lambda_b^0\to \Lambda^0 E_{\mathrm{miss}}$ decays under three different new physics hypotheses: the heavy new particles, the light neutral vector particles, and the axion-like particles. We find that all three hypotheses can resolve the Belle-II excess, and they can be clearly distinguished by the longitudinal polarization fraction of $K^*$. Furthermore, we discover that the longitudinal polarization fraction of $\Lambda$ can be used to distinguish the chirality of the effective operators.  
\end{abstract}
\maketitle

\section{Introduction}
\label{sec:introduction}

The rare $b \to s \nu \bar{\nu}$ transitions, as flavour-changing neutral-current (FCNC) processes, do not occur at the tree level and are highly suppressed by the Glashow-Iliopoulos-Maiani (GIM) mechanism~\cite{Glashow:1970gm} at higher orders within the Standard Model (SM). Compared to the semileptonic decay into a pair of charged leptons, the theoretical predictions for these observables are cleaner due to the absence of long-distance contributions from $c \bar{c}$ resonances~\cite{Ciuchini:2015qxb}. Based on these, the decays caused by $b \to s \nu \bar{\nu}$ play an important role in testing the SM.

Very recently, the Belle-II collaboration reported on the evidence for $B^+ \to K^+ E_{\mathrm{miss}}$ decay, with a branching ratio~\cite{Belle-II:2023esi}
\begin{align}
\label{eq:BKexp}
\mathcal{B}(B^+ \to K^+ E_{\mathrm{miss}})_{\mathrm{exp}} = (23 \pm 7) \times 10^{-6}.
\end{align}
In the SM, the missing energy $E_{\mathrm{miss}}$ is carried by a pair of massless neutrinos. Using the vector form factor $f_+(q^2)$ provided in Ref.~\cite{Gubernari:2023puw}, which is derived based on the analysis results from lattice Quantum Chromodynamics (LQCD)~\cite{Bouchard:2013eph,Bailey:2015dka,Parrott:2022rgu} and dispersive bounds, we obtain the SM prediction for the $B^+ \to K^+ \nu \bar{\nu}$ decay as follows
\begin{align}
\label{eq:BKsm}
\mathcal{B}(B^+ \to K^+ \nu \bar{\nu})_{\mathrm{SM}} = (5.09 \pm 0.41) \times 10^{-6}.
\end{align}
This result has removed the tree-level long-distance contribution from $B^+ \to \tau^+ \nu$ with $\tau^+ \to K^+ \bar{\nu}$~\cite{Kamenik:2009kc}. It agrees well with some recent predicted values in the SM (such as those in Refs.~\cite{Parrott:2022zte,Becirevic:2023aov,Tian:2024ubt}), but there is $2.6\sigma$ discrepancy from the above Belle-II result.

There are already many new physics (NP) models aiming to resolve the Belle-II excess~\cite{Athron:2023hmz,Bause:2023mfe,Allwicher:2023xba,Abdughani:2023dlr,Chen:2023wpb,He:2023bnk,Datta:2023iln,Altmannshofer:2023hkn,McKeen:2023uzo,Fridell:2023ssf,Ho:2024cwk,Chen:2024jlj,Gabrielli:2024wys,Hou:2024vyw,Chen:2024cll,He:2024iju,Bolton:2024egx,Marzocca:2024hua,Rosauro-Alcaraz:2024mvx,Eguren:2024oov,Buras:2024ewl,Hati:2024ppg,Allwicher:2024ncl,Becirevic:2024iyi,Altmannshofer:2024kxb,Buras:2024mnq}. In addition to postulating the existence of heavy NP as a solution, the authors in Ref.~\cite{Altmannshofer:2023hkn} provide a different approach, suggesting that this anomaly can be reinterpreted through the search for two-body $B \to K X$ decays, assuming that the undetectable particle $X$ is stable or its decay is invisible. They pointed out that this anomaly exhibits a localized characteristic. After taking $B \to K X$ into consideration and fitting the kinematic distribution, they determined that the mass of light NP particle $X$ is around 2 GeV, with a significance of approximately 3.6 $\sigma$. In comparison, the deviation observed by Belle-II under the assumption of the existence of only heavy NP particle was 2.6 $\sigma$, significantly lower than the former. This conclusion has been confirmed by subsequent works~\cite{Fridell:2023ssf,Bolton:2024egx}. For example, the authors in Ref.~\cite{Bolton:2024egx} considered the NP scenarios in which there are up to two new light invisible particles (with spin ranging from $0$ to $3/2$) in the final state. They found that in two-body decay, the value $m_X = (2.1 \pm 0.1)$ GeV provides the best fit to the data, with a significance of $4.5\sigma$ over the SM. Additionally, they pointed out that two-body decay kinematics seems to give a better fit to Belle-II data than three-body decay spectra.

In this work, we will consider three different NP hypotheses beyond the SM. They are: 1) the existence of only heavy new particles, with all observed missing energy always being carried by SM neutrinos; 2) the existence of only light neutral vector particles $Z'$, with the excess missing energy observed in experiment always being carried by a single $Z'$; 3) the existence of only light pseudoscalar or axion-like particles $a$, with the excess missing energy observed in experiment always being carried by a single $a$. For the first hypothesis, the heavy NP particles can only affect $b\to s \nu \bar{\nu}$ through off-shell intermediate states, and the corresponding contributions are encoded in the Wilson coefficients corresponding to two six-dimensional effective operators. In calculations, these two effective operators can usually be decomposed into two parts: $b\to s V^*$ and $V^* \to \nu \bar{\nu}$, where the virtual vector boson $V^*$ includes both spin-0 and spin-1 states. The contributions of particles $V^*$, $Z'$, and $a$, which carry different spin information, to the $b \to s E_{\mathrm{miss}}$ decay are highly dependent on the spin quantum numbers of the initial and final hadrons. Therefore, we will discuss $0^- \to 0^-$, $0^- \to 1^-$, and $1/2^+ \to 1/2^+$ decays respectively. Specifically, we will discuss the contributions of the aforementioned three different NP hypotheses to the $B^+\to K^+ E_{\mathrm{miss}}$, $B^0\to K^{*0} E_{\mathrm{miss}}$, and $\Lambda_b^0\to \Lambda^0 E_{\mathrm{miss}}$ decays one by one, and use the observables of these processes to present schemes for distinguishing different NP scenarios.

Our paper is organized as follows. In Sec.~\ref{sec:models}, we present the NP models and the analytical expressions for the contributions of NP to the $B^+\to K^+ E_{\mathrm{miss}}$, $B^0\to K^{*0} E_{\mathrm{miss}}$, and $\Lambda_b^0\to \Lambda^0 E_{\mathrm{miss}}$ decays. In Sec.~\ref{sec:numerical}, we provide our numerical results and discussions. Our conclusions are made in Sec.~\ref{sec:conclusions}.

\section{Models and observables}
\label{sec:models}

This section introduces the NP hypotheses to be considered, as well as their contributions to the $B^+\to K^+ E_{\mathrm{miss}}$, $B^0\to K^{*0} E_{\mathrm{miss}}$, and $\Lambda_b^0\to \Lambda^0 E_{\mathrm{miss}}$ decays.

\subsection{Heavy new particles}
\label{subsec:NPH}
Assuming that all NP beyond the SM are heavy, with their masses much greater than the electroweak scale. After integrating out the heavy NP particles and the heavy particles in the SM, namely the top quark, the $W^\pm$, $Z^0$ and Higgs boson, we can obtain the low-energy effective Hamiltonian suitable for describing the $b\to s \nu \bar{\nu}$ transitions~\cite{Buras:2014fpa}
\begin{align}
\label{eq:NPH}
\mathcal{H}_\mathrm{eff} = - \frac{4 G_F}{\sqrt{2}}  \lambda_t
\left(C_L \mathcal{O}_L + C_R \mathcal{O}_R \right)
+ \mathrm{H.c.} ,
\end{align}
with $\lambda_t = V_{tb} V^*_{ts}$ and 
\begin{align}
\mathcal{O}_L &= \frac{\alpha}{4 \pi} \left(\overline{s} \gamma_\mu P_L b \right) \left(\overline{\nu} \gamma^\mu \left(1 - \gamma_5 \right) \nu \right) ,
\\
\mathcal{O}_R &= \frac{\alpha}{4 \pi} \left(\overline{s} \gamma_\mu P_R b \right) \left(\overline{\nu} \gamma^\mu \left(1 - \gamma_5 \right) \nu \right) .
\end{align}
Here, $G_F$ is the Fermi constant, $V_{tb}$ and $V_{ts}$ are the Cabibbo-Kobayashi-Maskawa (CKM) matrix entries, $\alpha$ is the fine-structure constant, and the chirality projectors $P_{L,R} = (1 \mp \gamma_5)/2$. In the SM, $C_L = C_L^\mathrm{SM}$ and $C_R = 0$, where $C_L^\mathrm{SM} = -6.32 \pm 0.07$~\cite{Becirevic:2023aov} includes the next-to-leading order (NLO) Quantum Chromodynamics (QCD) corrections~\cite{Buchalla:1993bv,Misiak:1999yg,Buchalla:1998ba} and the two-loop electroweak contributions~\cite{Brod:2010hi}.

In this hypothesis, the differential decay rate of $B \to K \nu \bar{\nu}$ is given by 
\begin{align}
\label{eq:NPHK}
\frac{d\Gamma (B \to K \nu \bar{\nu})}{dq^2} = \frac{ G_F^2 |\lambda _t|^2 \alpha^2 \lambda _K^{3/2}(q^2)}{256 \pi ^5 m_B^3} |C_L+C_R|^2 f_+^2,
\end{align}
where $\lambda_K(q^2) \equiv \lambda(m_B^2,m_K^2,q^2)$.  The K{\"a}ll\'en function $\lambda(a,b,c) \equiv a^2 + b^2 + c^2 -2ab - 2ac - 2bc$. The $f_+(q^2)$ is a form factor of $B\to K$ and can be found in Ref.~\cite{Gubernari:2023puw}.

The differential decay rate of $B \to K^* \nu \bar{\nu}$ is given by 
\begin{align}
&\frac{d\Gamma^L}{dq^2} = \frac{G_F^2 |\lambda _t|^2 \alpha ^2 m_{K^*}^2 \sqrt{\lambda _{K^*}(q^2)}}{4 \pi ^5 m_B} |C_R-C_L|^2 A_{12}^2,\\
&\frac{d\Gamma^T}{dq^2} = \frac{G_F^2 |\lambda _t|^2 \alpha ^2 q^2 \sqrt{\lambda _{K^*}(q^2)} }{128 \pi ^5 m_B^3}\times \nonumber \\ 
&\qquad\qquad\Bigg[ \left(m_B+m_{K^*}\right)^2 |C_R-C_L|^2 A_1^2 \nonumber \\ 
&\qquad\qquad\quad+ \frac{\lambda _{K^*}(q^2)}{\left(m_B+m_{K^*}\right)^2} |C_L+C_R|^2 V^2 \Bigg], \\
&\frac{d\Gamma (B \to K^* \nu \bar{\nu})}{dq^2} =  \frac{d\Gamma^L}{dq^2} + \frac{d\Gamma^T}{dq^2}, \label{eq:GamKs}
\end{align}
where $\lambda_{K^*}(q^2) \equiv \lambda(m_B^2,m_{K^*}^2,q^2)$. Here, $d\Gamma^L/dq^2$ and $d\Gamma^T/dq^2$ represent the differential decay rates for the longitudinal and transverse polarization of the vector meson $K^*$, respectively. The $A_{12}(q^2)$, $A_1(q^2)$ and $V(q^2)$ are form factors of $B\to K^*$ and can be found in Ref.~\cite{Gubernari:2023puw}. In addition to the decay rate, we also consider an additional observable, that is, the longitudinal polarization fraction of $K^*$, which is defined as follows
\begin{align}
P_L^{K^*} = \frac{\int_{0}^{(m_B-m_{K^*})^2} \frac{d\Gamma^L}{dq^2} dq^2}{\int_{0}^{(m_B-m_{K^*})^2} \frac{d\Gamma}{dq^2} dq^2}.
\end{align}

The differential decay rate of $\Lambda_b \to \Lambda \nu \bar{\nu}$ is given by 
\begin{align}
	&\frac{d\Gamma^\pm}{dq^2} = \frac{G_F^2 |\lambda _t|^2 \alpha^2 \sqrt{\lambda _{\Lambda }(q^2)}}{512 \pi ^5 m_{\Lambda _b}^3} \times \nonumber\\
	&\Big[\Big| \left(m_{\Lambda_b}+m_{\Lambda}\right)\sqrt{s_-(q^2)} (C_L+C_R) F_+ \nonumber\\
	&\pm 
	\left(m_{\Lambda _b} - m_{\Lambda }\right) \sqrt{s_+(q^2)} (C_R-C_L) G_+ \Big|^2 +  2 q^2 \times \nonumber \\
	& \left| \sqrt{s_-(q^2)} (C_L+C_R) F_\perp \pm \sqrt{s_+(q^2)} (C_R-C_L) G_\perp
	\right|^2\Big],\\
	&\frac{d\Gamma (\Lambda_b \to \Lambda \nu \bar{\nu})}{dq^2} =\frac{d\Gamma^-}{dq^2} + \frac{d\Gamma^+}{dq^2}\nonumber \\
	 &=\frac{G_F^2 \left| \lambda _t\right|^2 \alpha ^2 \sqrt{\lambda _{\Lambda }(q^2)} }{256 \pi ^5 m_{\Lambda _b}^3} \times \nonumber\\
	 &\Big[ s_-(q^2) |C_L+C_R|^2
	\left( \left(m_{\Lambda _b}+m_{\Lambda }\right)^2 F_+^2 +2 q^2 F_\perp^2\right) + \nonumber\\
	&  s_+(q^2) |C_R-C_L|^2 \left( \left(m_{\Lambda _b}-m_{\Lambda }\right)^2 G_+^2 +2 q^2 G_\perp^2
	\right)\Big], \label{eq:GamL}
\end{align}
where $\lambda_{\Lambda}(q^2) \equiv \lambda(m_{\Lambda _b}^2,m_{\Lambda }^2,q^2)$ and $s_\pm(q^2) \equiv (m_{\Lambda _b} \pm m_{\Lambda })^2 - q^2$, as well as $\lambda_{\Lambda}(q^2) = s_+(q^2) s_-(q^2)$. Here, $d\Gamma^-/dq^2$ and $d\Gamma^+/dq^2$ represent the differential decay rates when the helicity of the baryon $\Lambda$ are $-1/2$ and $+1/2$ respectively. The $F_{+,\perp}(q^2)$ and $G_{+,\perp}(q^2)$ are form factors of $\Lambda_b \to \Lambda$ and can be found in Ref.~\cite{Detmold:2016pkz}.\footnote{To avoid confusion with some notations in this work, we have respectively changed the lowercase letters $f_{0,+,\perp}$ and $g_{0,+,\perp}$ in Ref.~\cite{Detmold:2016pkz} into the uppercase letters $F_{0,+,\perp}$ and $G_{0,+,\perp}$.} Besides the decay rate, we also include the longitudinal polarization fraction
\begin{align}
P_L^\Lambda = \frac{\int_{0}^{(m_{\Lambda_b}-m_\Lambda)^2} \frac{d\Gamma^-}{dq^2} - \frac{d\Gamma^+}{dq^2} dq^2}{\int_{0}^{(m_{\Lambda_b}-m_\Lambda)^2} \frac{d\Gamma^-}{dq^2} + \frac{d\Gamma^+}{dq^2} dq^2}.
\end{align}

\subsection{Light neutral vector particles}
\label{subsec:NPZ}
Assuming that the Belle-II excess originates from the light neutral vector particles $Z'$, which are stable or their decays are invisible, manifested through the decay $B \to K Z'$. Considering local operators up to dimension six, the effective Lagrangian is~\cite{Altmannshofer:2017bsz,Altmannshofer:2023hkn}
\begin{align}
	\mathcal{L}_{Z'} = \Bigg[ & g_L^{(4)\prime} Z'_\mu (\bar{s} \gamma^\mu P_L b)  +  \frac{g_L^{(5)}}{\Lambda} Z'_{\mu\nu} (\bar{s} \sigma^{\mu\nu} P_R b)  \nonumber\\
	&+  \frac{g_L^{(6)\prime}}{\Lambda^2} \partial^\nu Z'_{\mu\nu} (\bar{s} \gamma^\mu P_L b)  +  \mathrm{h.c.} \Bigg] + \{L \leftrightarrow R\},
\end{align}
with the $Z'$ field strength tensor $Z'_{\mu\nu} = \partial_\mu Z'_\nu - \partial_\nu Z'_\mu$. For the convenience of later discussions, we introduce the notations $g^{(4)}_{L,R} = g^{(4)\prime}_{L,R} + (m_{Z'}^2/\Lambda^2) g^{(6)\prime}_{L,R}$, as well as the vector couplings $g_V^{(d)} = g_R^{(d)} + g_L^{(d)}$ and axial-vector couplings $g_A^{(d)} = g_R^{(d)} - g_L^{(d)}$.

In this hypothesis, the decay rate of $B \to K Z'$ is 
\begin{align}
\label{eq:NPKZp}
\Gamma(B \to K Z') = \frac{\lambda _{K}^{3/2}(m_{Z'}^2)}{64 \pi  m_B^3 m_{Z'}^2}  \left|g^{(4)}_V f_+
+ \frac{2m_{Z'}^2 g^{(5)}_V f_T}{\left(m_B+m_K\right)\Lambda }\right|^2,
\end{align}
where the $B\to K$ form factors $f_+(m_{Z'}^2)$ and $f_T(m_{Z'}^2)$ can be found in Ref.~\cite{Gubernari:2023puw}. The decay rate of $B\to K E_{\mathrm{miss}}$, which is measured in the Belle-II experiment, can be expressed as
\begin{align}
\Gamma(B\to K E_{\mathrm{miss}})_{Z'} = \Gamma(B \to K \nu \bar{\nu})_{\mathrm{SM}} + \Gamma(B \to K Z').
\end{align}

The decay rate of $B \to K^* Z'$ is 
\begin{align}
	&\Gamma^L_{Z'} = \frac{m_{K^*}^2 \sqrt{\lambda _{K^*}(m_{Z'}^2)}}{\pi  m_B m_{Z'}^2} \left| g_A^{(4)} A_{12} + i \frac{m_{Z'}^2 g_A^{(5)} T_{23} }{ 
		\left(m_B+m_{K^*}\right)\Lambda}\right|^2,\\
	&\Gamma^T_{Z'} = \frac{\sqrt{\lambda _{K^*}(m_{Z'}^2)} }{32 \pi  m_B^3} 
	\Bigg[\left(m_B+m_{K^*}\right)^2 \nonumber\\
	&\qquad\qquad \times \left| g_A^{(4)} A_1+ i \frac{2
		\left(m_B-m_{K^*}\right) g_A^{(5)} T_2}{\Lambda }\right|^2 \nonumber\\
	&\qquad\qquad +\lambda_{K^*}(m_{Z'}^2) \left| \frac{g_V^{(4)}
		V}{m_B+m_{K^*}}- i \frac{2 g_V^{(5)} T_1}{\Lambda }\right|^2\Bigg],\\
&\Gamma(B \to K^* Z') = \Gamma^L_{Z'} + \Gamma^T_{Z'}, \label{eq:GamKsZp}
\end{align}
where the $B\to K^*$ form factors $A_{12}(m_{Z'}^2)$, $T_{23}(m_{Z'}^2)$, $A_1(m_{Z'}^2)$, $T_2(m_{Z'}^2)$, $V(m_{Z'}^2)$ and $T_1(m_{Z'}^2)$ can be found in Ref.~\cite{Gubernari:2023puw}. Similarly, the observable $\Gamma(B\to K^* E_{\mathrm{miss}})$ can be obtained by adding $\Gamma(B \to K^* \nu \bar{\nu})_{\mathrm{SM}}$ to the above Eq.~\eqref{eq:GamKsZp}. The longitudinal $K^*$ polarization fraction is given by
\begin{align}
	P_{L Z'}^{K^*} = \frac{\int_{0}^{(m_B-m_{K^*})^2} \left(\frac{d\Gamma^L}{dq^2}\right)_\mathrm{SM} dq^2 + \Gamma^L_{Z'}}{\Gamma(B\to K^* E_{\mathrm{miss}})_{Z'}}.
\end{align}

The decay rate of $\Lambda_b \to \Lambda Z'$ is 
\begin{align}
	\Gamma^\pm_{Z'} = \frac{\sqrt{\lambda _{\Lambda}(m_{Z'}^2)}}{32 \pi  m_{\Lambda _b}^3} \Big(&\left| x_1 \pm x_2+x_3 \pm x_4\right|^2 \nonumber\\
	&+\left| x_5 \pm x_6+x_7 \pm x_8\right|^2\Big),\\
	\Gamma(\Lambda_b \to \Lambda Z') = \Gamma^-_{Z'} &+ \Gamma^+_{Z'}  \nonumber\\
	=\frac{\sqrt{\lambda _{\Lambda}(m_{Z'}^2)}}{16 \pi  m_{\Lambda _b}^3} \Big(&\left| x_1+x_3\right|^2+\left| x_2+x_4\right|^2 \nonumber\\
	&+\left| x_5+x_7\right|
	^2+\left| x_6+x_8\right|^2\Big), \label{eq:GamLZp}
\end{align}
with
\begin{align}
	x_1 &\equiv \frac{\left(m_{\Lambda _b}+m_{\Lambda }\right) \sqrt{s_-(m_{Z'}^2)}}{2 m_{Z'}} g_V^{(4)} F_+ ,\\
	x_2 &\equiv \frac{ \left(m_{\Lambda _b}-m_{\Lambda }\right) \sqrt{s_+(m_{Z'}^2)} }{2 m_{Z'}} g_A^{(4)} G_+ , \\
	x_3 &\equiv \frac{m_{Z'} \sqrt{s_-(m_{Z'}^2)} }{\Lambda } g_V^{(5)} h_+, \\
	x_4 &\equiv \frac{m_{Z'} \sqrt{s_+(m_{Z'}^2)} }{\Lambda } g_A^{(5)} \tilde{h}_+, \\
	x_5 &\equiv \sqrt{\frac{s_-(m_{Z'}^2)}{2}} g_V^{(4)} F_\perp, \\
	x_6 &\equiv \sqrt{\frac{s_+(m_{Z'}^2)}{2}} g_A^{(4)} G_\perp, \\
	x_7 &\equiv \frac{\left(m_{\Lambda _b}+m_{\Lambda }\right) \sqrt{2s_-(m_{Z'}^2)} }{\Lambda } g_V^{(5)} h_\perp, \\
	x_8 &\equiv \frac{ \left(m_{\Lambda _b}-m_{\Lambda }\right)\sqrt{2s_+(m_{Z'}^2)}}{\Lambda } g_A^{(5)} \tilde{h}_\perp.
\end{align}
Here the $\Lambda_b \to \Lambda$ form factors $F_{+,\perp}(m_{Z'}^2)$, $G_{+,\perp}(m_{Z'}^2)$, $h_{+,\perp}(m_{Z'}^2)$ and $\tilde{h}_{+,\perp}(m_{Z'}^2)$ can be found in Ref.~\cite{Detmold:2016pkz}. The observable $\Gamma(\Lambda_b\to \Lambda E_{\mathrm{miss}})$ can be obtained by adding $\Gamma(\Lambda_b \to \Lambda \nu \bar{\nu})_{\mathrm{SM}}$ to the above Eq.~\eqref{eq:GamLZp}. The longitudinal polarization fraction is
\begin{align}
	P_{L Z'}^{\Lambda} = \frac{\int_{0}^{(m_{\Lambda_b}-m_\Lambda)^2} \left(\frac{d\Gamma^-}{dq^2} - \frac{d\Gamma^+}{dq^2}\right)_\mathrm{SM} dq^2 + \Gamma^-_{Z'} - \Gamma^+_{Z'}}{\Gamma(\Lambda_b\to \Lambda E_{\mathrm{miss}})_{Z'}}.
\end{align}

\subsection{Axion-like particles}
\label{subsec:NPa}
Assuming that the Belle-II excess originates from the massive pseudoscalar or axion-like particles (ALPs) $a$, which are stable or their decays are invisible, manifested through the decay $B \to K a$. The corresponding effective Lagrangian can be expressed either by coupling the derivative of field $a$ with the $bs$ (axial-)vector current or by directly coupling field $a$ with the $bs$ (pseudo-)scalar current, and these two representations are equivalent up to total derivatives.
\begin{align}
\mathcal{L}_a &= \frac{\partial_\mu a}{f} \left(\kappa_L \bar{s} \gamma^\mu P_L b + \kappa_R \bar{s} \gamma^\mu P_R b\right) + \mathrm{h.c.}\\
&= i \frac{a}{2f}\Big[(m_b - m_s)(\kappa_L + \kappa_R)\bar{s} b \nonumber\\
&\qquad\quad  + (m_b + m_s)(\kappa_L - \kappa_R)\bar{s} \gamma_5 b \Big]+ \mathrm{h.c.},
\end{align}
where $f$ is the decay constant of $a$.

In this hypothesis, the decay rate of $B \to K a$ is
\begin{align}
	\Gamma(B \to K a) = \frac{\left(m_B^2-m_K^2\right)^2 \sqrt{\lambda _{K}(m_a^2)} }{64 \pi  f^2 m_B^3} |\kappa_L + \kappa_R|^2  f_0^2 ,
\end{align}
where the $B \to K$ form factor $f_0(m_a^2)$, which did not enter the Eqs.~\eqref{eq:NPHK} and \eqref{eq:NPKZp}, can be found in Ref.~\cite{Gubernari:2023puw}. The decay rate of $B\to K E_{\mathrm{miss}}$ has now changed to
\begin{align}
	\Gamma(B\to K E_{\mathrm{miss}})_a = \Gamma(B \to K \nu \bar{\nu})_{\mathrm{SM}} + \Gamma(B \to K a).
\end{align}

The decay rate of $B \to K^* a$ is 
\begin{align}
	\Gamma(B \to K^* a) = \Gamma^L_a = \frac{\lambda _{K^*}^{3/2}(m_a^2)}{64 \pi  f^2 m_B^3} |\kappa_R-\kappa_L|^2 A_0^2,
\end{align}
where the $B \to K^*$ form factor $A_0(m_a^2)$, which did not enter the Eqs.~\eqref{eq:GamKs} and \eqref{eq:GamKsZp}, can be found in Ref.~\cite{Gubernari:2023puw}. The contributions of ALPs $a$ are entirely on the longitudinal $K^*$ polarization. The missing energy is carried by SM neutrinos and ALPs, and the observable $\Gamma(B\to K^* E_{\mathrm{miss}})$ is determined by $\Gamma(B \to K^* \nu \bar{\nu})_{\mathrm{SM}} + \Gamma(B \to K^* a)$ at this time. The longitudinal $K^*$ polarization fraction now becomes
\begin{align}
	P_{L a}^{K^*} = \frac{\int_{0}^{(m_B-m_{K^*})^2} \left(\frac{d\Gamma^L}{dq^2}\right)_\mathrm{SM} dq^2 + \Gamma^L_{a}}{\Gamma(B\to K^* E_{\mathrm{miss}})_{a}}.
\end{align}

The decay rate of $\Lambda_b \to \Lambda a$ is 

\begin{align}
	&\Gamma^\pm_a = \frac{\sqrt{\lambda _{\Lambda}(m_a^2)}}{128 \pi  f^2 m_{\Lambda _b}^3} \Big| \left(m_{\Lambda _b}-m_{\Lambda }\right) \sqrt{s_+(m_a^2)} (\kappa_L + \kappa_R) F_0 \nonumber\\
	 &\qquad\qquad \pm \left(m_{\Lambda _b
	}+m_{\Lambda}\right) \sqrt{s_-(m_a^2)} (\kappa_R - \kappa_L) G_0\Big|^2, \\
	&\Gamma(\Lambda_b \to \Lambda a) = \Gamma^-_a + \Gamma^+_a \nonumber\\
	&\quad = \frac{\sqrt{\lambda _{\Lambda}(m_a^2)}}{64 \pi  f^2 m_{\Lambda _b}^3} \Big[ \left(m_{\Lambda _b}-m_{\Lambda }\right)^2 s_+(m_a^2) \left|\kappa_L + \kappa_R \right|^2 F_0^2 \nonumber\\
	&\qquad\qquad+ \left(m_{\Lambda _b
	}+m_{\Lambda}\right)^2 s_-(m_a^2) \left| \kappa_R-\kappa_L \right|^2 G_0^2  \Big],
\end{align}
where the $\Lambda_b \to \Lambda$ form factors $F_0(m_a^2)$ and $G_0(m_a^2)$, which did not enter the Eqs.~\eqref{eq:GamL} and \eqref{eq:GamLZp}, can be found in Ref.~\cite{Detmold:2016pkz}. Similar to the above, the result of the decay rate of $\Gamma(\Lambda_b\to \Lambda E_{\mathrm{miss}})$ is now $\Gamma(\Lambda_b \to \Lambda \nu \bar{\nu})_{\mathrm{SM}} +\Gamma(\Lambda_b \to \Lambda a)$, and the longitudinal polarization fraction has changed to 
\begin{align}
	P_{L a}^{\Lambda} = \frac{\int_{0}^{(m_{\Lambda_b}-m_\Lambda)^2} \left(\frac{d\Gamma^-}{dq^2} - \frac{d\Gamma^+}{dq^2}\right)_\mathrm{SM} dq^2 + \Gamma^-_{a} - \Gamma^+_{a}}{\Gamma(\Lambda_b\to \Lambda E_{\mathrm{miss}})_{a}}.
\end{align}

\section{Numerical results and discussions}
\label{sec:numerical}

\begin{table}[t]
\tabcolsep 0.12in
\renewcommand\arraystretch{1.2}
\begin{center}
\caption{\label{tab:input} \small Summary of input parameters used throughout this paper. }
\vspace{0.18cm}
\begin{tabular}{ccc} 
\hline
Parameter& Value & References
\\ \hline
$G_F$ & $1.1663788(6)\times 10^{-5}$ GeV$^{-2}$  &  \cite{ParticleDataGroup:2024cfk}
\\
$\alpha$ & $1/128$   &  \cite{ParticleDataGroup:2024cfk}
\\
$m_K$ & $493.677(15) \times 10^{-3}$ GeV  &  \cite{ParticleDataGroup:2024cfk}
\\
$m_{K^*}$ & $895.55(20) \times 10^{-3}$ GeV  &  \cite{ParticleDataGroup:2024cfk}
\\
$m_{B^+}$ & $5279.41(7) \times 10^{-3}$ GeV  &  \cite{ParticleDataGroup:2024cfk}
\\
$m_{B^0}$ & $5279.72(8) \times 10^{-3}$ GeV  &  \cite{ParticleDataGroup:2024cfk}
\\
$m_{\Lambda}$ & $1115.683(6) \times 10^{-3}$ GeV  &  \cite{ParticleDataGroup:2024cfk}
\\
$m_{\Lambda_b}$ & $5619.60(17) \times 10^{-3}$ GeV  &  \cite{ParticleDataGroup:2024cfk}
\\
$\tau_{B^+}$ & $1.638(4)$ ps  &  \cite{ParticleDataGroup:2024cfk}
\\
$\tau_{B^0}$ & $1.517(4)$ ps  &  \cite{ParticleDataGroup:2024cfk}
\\
$\tau_{\Lambda_b}$ & $1.471(9)$ ps  &  \cite{ParticleDataGroup:2024cfk}
\\
$|V_{tb}|$ & $1.010(27)$   &  \cite{ParticleDataGroup:2024cfk}
\\
$|V_{ts}|$ & $41.5(9) \times 10^{-3}$   &  \cite{ParticleDataGroup:2024cfk}
\\
$C_L^\mathrm{SM}$ & $-6.32(7)$   &  \cite{Becirevic:2023aov}
\\
\hline
\multicolumn{2}{c}{$B \to K$ form factors} & \cite{Parrott:2022rgu,Bailey:2015dka,Bouchard:2013eph,Gubernari:2023puw}
\\
\multicolumn{2}{c}{$B \to K^*$ form factors} & \cite{Gubernari:2018wyi,Horgan:2013hoa,Horgan:2015vla,Gubernari:2023puw}
\\
\multicolumn{2}{c}{$\Lambda_b \to \Lambda$ form factors} & \cite{Detmold:2016pkz,Blake:2022vfl}
\\
\hline
\end{tabular}
\end{center}
\end{table}

All the theoretical input parameters required in this work are summarized in Tab.~\ref{tab:input}. By using them, we can obtain the following predicted values within the SM.
\begin{align}
\mathcal{B}(B^+ \to K^+ \nu \bar{\nu})_\mathrm{SM} &= (5.09 \pm 0.41) \times 10^{-6},\\
\mathcal{B}(B^0 \to K^{*0} \nu \bar{\nu})_\mathrm{SM} &= (8.79 \pm 1.05) \times 10^{-6},\\
P^{K^*}_{L \mathrm{SM}} &= 0.44 \pm 0.02,\\
\mathcal{B}(\Lambda_b \to \Lambda \nu \bar{\nu})_\mathrm{SM} &= (8.39 \pm 1.15) \times 10^{-6},\\
P^{\Lambda}_{L \mathrm{SM}} &= 0.93 \pm 0.02.
\end{align}
The above results indicate that, within the SM, the branching ratios of the exclusive $b \to s \nu \bar{\nu}$ processes are approximately on the order of $10^{-6}$. Among the final state of $B^0 \to K^{*0} \nu \bar{\nu}$ decay, the longitudinal $K^*$ accounts for about 44\%, while in the final state of $\Lambda_b \to \Lambda \nu \bar{\nu}$ decay, the majority are baryon $\Lambda$ with helicity $-1/2$, accounting for approximately 96\%. To be conservative, we have retained the uncertainties arising from all input parameters, with the uncertainties from irrelevant parameters being summed in quadrature.

Beyond the SM, we consider the following eight NP scenarios.\\
\textbf{SH1}: heavy NP particles only contributes to a non-zero $C_L^\mathrm{NP}$;\\
\textbf{SH2}: heavy NP particles only contributes to a non-zero $C_R$;\\
\textbf{SZ1}: light vectors only contributes to a non-zero $g^{(4)}_L$, which can be realized through non-zero $g^{(4)\prime}_L$ and/or $g^{(6)\prime}_L$;\\
\textbf{SZ2}: light vectors only contributes to a non-zero $g^{(4)}_R$, which can be realized through non-zero $g^{(4)\prime}_R$ and/or $g^{(6)\prime}_R$;\\
\textbf{SZ3}: light vectors only contributes to a non-zero $g^{(5)}_L$;\\
\textbf{SZ4}: light vectors only contributes to a non-zero $g^{(5)}_R$;\\
\textbf{Sa1}: light ALPs only contributes to a non-zero $\kappa_L$;\\
\textbf{Sa2}: light ALPs only contributes to a non-zero $\kappa_R$.\\
Above, we only consider the scenario where the NP particles contribute to an operator of a single chirality.

Next, we will discuss the impacts of NP on the observables $\mathcal{B}(B^+ \to K^+ E_{\mathrm{miss}})$, $\mathcal{B}(B^0 \to K^{*0} E_{\mathrm{miss}})$, $P^{K^*}_{L}$, $\mathcal{B}(\Lambda_b \to \Lambda E_{\mathrm{miss}})$, and $P^{\Lambda}_{L}$ within each of the above mentioned scenarios. Currently, apart from the branching ratio of $B \to K E_{\mathrm{miss}}$ decay, the only experimental information available is the upper limit of $\mathcal{B}(B \to K^* E_{\mathrm{miss}})$, provided by BaBar~\cite{BaBar:2013npw} and Belle~\cite{Belle:2017oht} at the 90\% confidence level, respectively.
\begin{align}
\mathcal{B}(B^0 \to K^{*0} E_{\mathrm{miss}})_\mathrm{BaBar} &< 93 \times 10^{-6}, \label{eq:BaBarKs}\\
\mathcal{B}(B^0 \to K^{*0} E_{\mathrm{miss}})_\mathrm{Belle} &< 27 \times 10^{-6}. \label{eq:BelleKs}
\end{align}

\begin{figure}[t]
	\centering
	\includegraphics[width=0.4\textwidth]{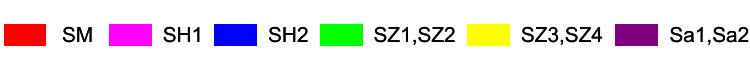}\\
	\includegraphics[width=0.4\textwidth]{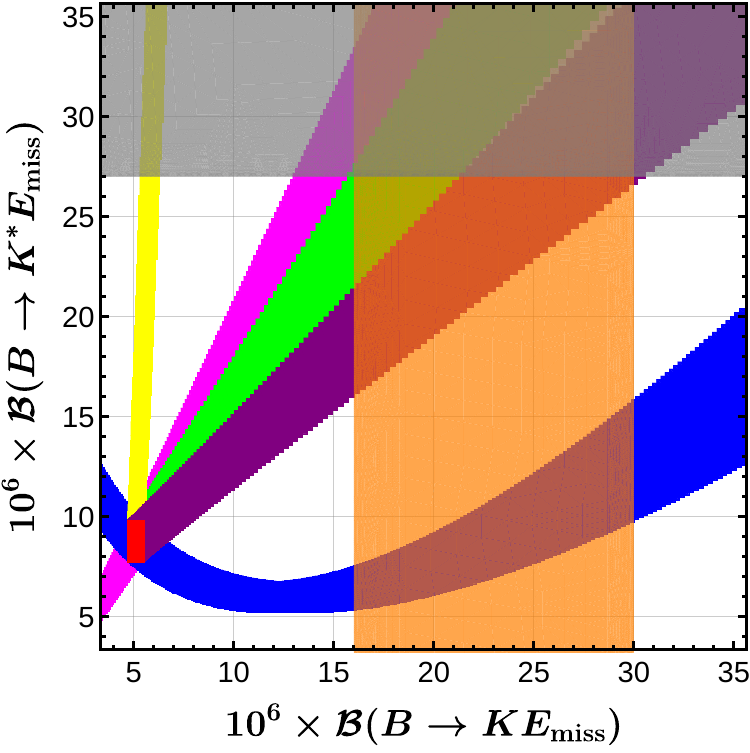}\\
	\includegraphics[width=0.4\textwidth]{leg1.pdf}\\
	\includegraphics[width=0.4\textwidth]{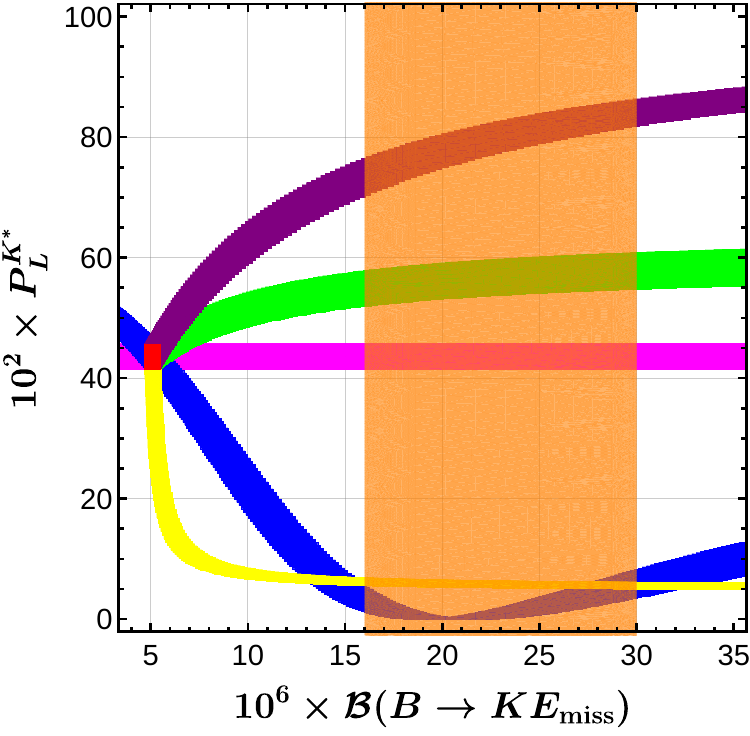}
	\caption{\label{fig:BKs}The figure displays the $\mathcal{B}(B^+ \to K^+ E_{\mathrm{miss}})-\mathcal{B}(B^0 \to K^{*0} E_{\mathrm{miss}})$ correlation (top) and $\mathcal{B}(B^+ \to K^+ E_{\mathrm{miss}})-P^{K^*}_{L}$ correlation (bottom) for different NP scenarios. The SM predictions are represented by red rectangles. The light gray region in the upper plot is excluded by the experimental constraint on $\mathcal{B}(B^0 \to K^{*0} E_{\mathrm{miss}})$ given in Eq.~\eqref{eq:BelleKs}, and the light orange regions indicate the present experimental range~\eqref{eq:BKexp} quoted by Belle-II.}
\end{figure} 

\begin{figure}[t]
	\centering
	\includegraphics[width=0.4\textwidth]{leg1.pdf}\\
	\includegraphics[width=0.4\textwidth]{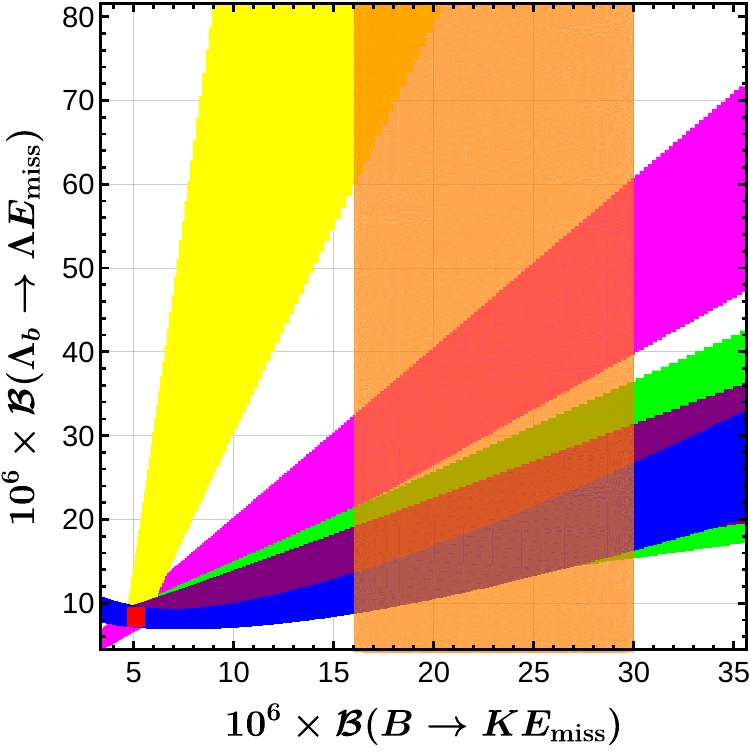}\\
	\includegraphics[width=0.4\textwidth]{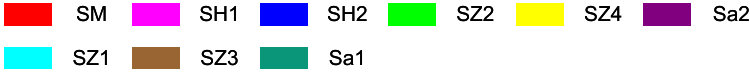}\\
	\includegraphics[width=0.4\textwidth]{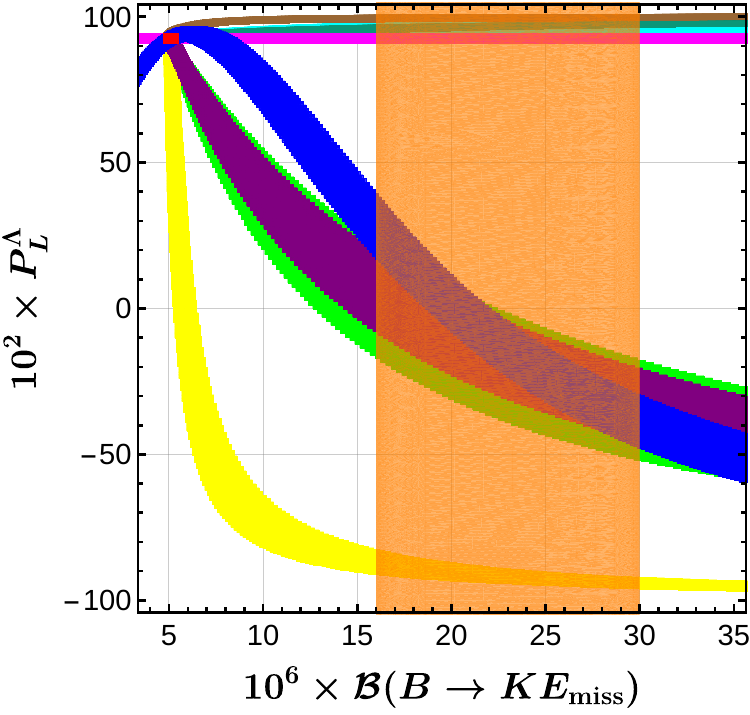}
	\caption{\label{fig:Lb2L}The figure displays the $\mathcal{B}(B^+ \to K^+ E_{\mathrm{miss}})-\mathcal{B}(\Lambda_b \to \Lambda E_{\mathrm{miss}})$ correlation (top) and $\mathcal{B}(B^+ \to K^+ E_{\mathrm{miss}})-P^{\Lambda}_{L}$ correlation (bottom) for different NP scenarios. The SM predictions are represented by red rectangles. The light orange regions indicate the present experimental range~\eqref{eq:BKexp} quoted by Belle-II.}
\end{figure} 

Within different NP scenarios, the correlation between the branching ratio of $B^+ \to K^+ E_{\mathrm{miss}}$ decay and that of $B^0 \to K^{*0} E_{\mathrm{miss}}$ decay, as well as the correlation between the branching ratio of $B^+ \to K^+ E_{\mathrm{miss}}$ decay and the longitudinal polarization fraction of $B^0 \to K^{*0} E_{\mathrm{miss}}$ decay, are presented in Fig.~\ref{fig:BKs}. We find that the results of scenarios \textbf{SZ1} and \textbf{SZ2} are exactly the same and cannot be distinguished. Similar situations occur between scenarios \textbf{SZ3} and \textbf{SZ4}, as well as between scenarios \textbf{Sa1} and \textbf{Sa2}. It is clearly visible that scenarios \textbf{SZ3} and \textbf{SZ4} (corresponding to the yellow region in the upper plot), which involve dimension-five left-handed and right-handed operators contributed by the light neutral vector particles, cannot enhance $\mathcal{B}(B^+ \to K^+ E_{\mathrm{miss}})$ to the Belle-II experimental region while satisfying the constraints given by Eq.~\eqref{eq:BelleKs} or even Eq.~\eqref{eq:BaBarKs}.

From the $\mathcal{B}(B^+ \to K^+ E_{\mathrm{miss}})-\mathcal{B}(B^0 \to K^{*0} E_{\mathrm{miss}})$ correlation plot, it can be obtained that all three NP hypotheses we considered have parameter spaces that can explain the Belle-II excess, especially for the right-handed operator contributed by the heavy NP particles (\textbf{SH2}) and the light ALPs (\textbf{Sa1} and \textbf{Sa2}), which possess larger parameter spaces. The $\mathcal{B}(B^0 \to K^{*0} E_{\mathrm{miss}})$ predicted by the scenario \textbf{SH2} is significantly smaller than that predicted by the other scenarios.

The question of whether the NP particles in $b \to s E_{\mathrm{miss}}$ are heavy or light can be clearly answered in the $\mathcal{B}(B^+ \to K^+ E_{\mathrm{miss}})-P^{K^*}_{L}$ correlation plot. Within the range that satisfies Eq.~\eqref{eq:BKexp}, the ALPs hypothesis can increase $P^{K^*}_{L}$ to approximately 80\% (\textbf{Sa1} and \textbf{Sa2}), the neutral vector particle hypothesis can elevate $P^{K^*}_{L}$ to around 55\% (\textbf{SZ1} and \textbf{SZ2}), while the heavy NP hypothesis can only keep $P^{K^*}_{L}$ at the value predicted by the SM (\textbf{SH1}) or reduce it to less than 10\% (\textbf{SH2}). Moreover, these regions are well-separated without any overlap. Scenarios \textbf{SZ3} and \textbf{SZ4} can also significantly reduce $P^{K^*}_{L}$, but they would simultaneously increase $\mathcal{B}(B^0 \to K^{*0} E_{\mathrm{miss}})$ to above $236\times 10^{-6}$, which clearly exceeds the upper limits provided by BaBar~\cite{BaBar:2013npw} and Belle~\cite{Belle:2017oht}.

In Fig.~\ref{fig:Lb2L}, we show the correlation between the branching ratio of $B^+ \to K^+ E_{\mathrm{miss}}$ decay and that of $\Lambda_b \to \Lambda E_{\mathrm{miss}}$ decay, as well as the correlation between the branching ratio of $B^+ \to K^+ E_{\mathrm{miss}}$ decay and the longitudinal polarization fraction of $\Lambda_b \to \Lambda E_{\mathrm{miss}}$ decay, in different NP scenarios. We also find that the results of scenarios \textbf{SZ1} and \textbf{SZ2} are exactly the same and cannot be distinguished in the $\mathcal{B}(B^+ \to K^+ E_{\mathrm{miss}})-\mathcal{B}(\Lambda_b \to \Lambda E_{\mathrm{miss}})$ correlation plot. Similar situations occur between scenarios \textbf{SZ3} and \textbf{SZ4}, as well as between scenarios \textbf{Sa1} and \textbf{Sa2}.

Unlike the previous three correlation plots, in the $\mathcal{B}(B^+ \to K^+ E_{\mathrm{miss}})-P^{\Lambda}_{L}$ correlation plot, the contributions from left-handed operators and right-handed operators in each NP hypothesis are completely separated. Specifically, in the presence of non-zero right-handed contributions (scenarios \textbf{SH2}, \textbf{SZ2}, \textbf{SZ4} and \textbf{Sa2}), $P^{\Lambda}_{L}$ rapidly decreases or even changes sign, indicating that there are fewer $\Lambda$ particles with helicity $-1/2$ than those with helicity $1/2$ in the decay products. On the other hand, in the presence of non-zero left-handed contributions (scenarios \textbf{SH1}, \textbf{SZ1}, \textbf{SZ3} and \textbf{Sa1}), $P^{\Lambda}_{L}$ increases slightly, and their predicted values are all within the range $P^\Lambda_{L \mathrm{SM}} \sim 1$ due to the constraint of the upper limit $1$. 

\section{Conclusions}
\label{sec:conclusions}
Recently, the $\mathcal{B}(B^+ \to K^+ E_{\mathrm{miss}})$ measurement released by the Belle-II collaboration is approximately $2.6\sigma$ higher than the SM prediction, which has sparked considerable research interest. Due to the obvious localized feature observed in the $q^2$ distribution published by the Belle-II, apart from explaining the Belle-II excess with heavy NP, another option is to consider light NP particles. 

In order to investigate whether the NP particles in the $b \to s E_{\mathrm{miss}}$ transitions are heavy or light, we study three exclusive processes involving hadrons with different spins, namely $B^+\to K^+ E_{\mathrm{miss}}$, $B^0\to K^{*0} E_{\mathrm{miss}}$, and $\Lambda_b^0\to \Lambda^0 E_{\mathrm{miss}}$ decays. In addition to their respective branching ratios, our research also includes the longitudinal polarization fractions $P_L^{K^*}$ of $B^0\to K^{*0} E_{\mathrm{miss}}$ and the $P_L^{\Lambda}$ of $\Lambda_b^0\to \Lambda^0 E_{\mathrm{miss}}$. We provide analytical expressions for the aforementioned five observables under three different NP hypotheses: the heavy new particles, the light neutral vector particles, and the ALPs.

We find that these three different NP hypotheses, as well as the chirality of the NP effects they provide, can be distinguished through correlation plots between the branching ratio of $B^+\to K^+ E_{\mathrm{miss}}$ decay and the other four observables, especially the $\mathcal{B}(B^+ \to K^+ E_{\mathrm{miss}})-P^{K^*}_{L}$ and $\mathcal{B}(B^+ \to K^+ E_{\mathrm{miss}})-P^{\Lambda}_{L}$ correlation plots. Under the experimental constraints of $\mathcal{B}(B^0\to K^{*0} E_{\mathrm{miss}})$, the heavy new particles, the light neutral vector particles, and the ALPs can all enhance $\mathcal{B}(B^+ \to K^+ E_{\mathrm{miss}})$ to within the measurement range, thereby explaining the excess observed by Belle-II. Meanwhile, the light neutral vector particles and the ALPs can increase $P^{K^*}_{L}$ to approximately 55\% and 80\% respectively (compared to about 44\% in the SM), while heavy NP particles either keep it unchanged (\textbf{SH1}) or reduce it to below 10\% (\textbf{SH2}). Due to the spin-half nature of $\Lambda_b$ and $\Lambda$ baryons, the contributions of left-handed and right-handed operators in the three different NP hypotheses to the longitudinal polarization fraction $P^{\Lambda}_{L}$ of $\Lambda_b^0\to \Lambda^0 E_{\mathrm{miss}}$ are entirely distinct. The $P^{\Lambda}_{L}$ can be used to distinguish the chirality of the effective operators. We anticipate more precise measurements of the aforementioned observables, particularly $P^{K^*}_{L}$ and $P^{\Lambda}_{L}$, from experiments such as Belle-II~\cite{Belle-II:2018jsg} and FCC-ee~\cite{Amhis:2023mpj}. This will help further deepen our understanding of the quark-level $b \to s E_{\mathrm{miss}}$ transitions.

\section*{Acknowledgements}

This work is supported by the National Natural Science Foundation of China under Grant No.~12105002, and the Guangxi Natural Science Foundation under
Grant No.~2023GXNSFBA026270.

\providecommand{\href}[2]{#2}\begingroup\raggedright\endgroup


\begin{thebibliography}{10}

\bibitem{Glashow:1970gm}
S.L.~Glashow, J.~Iliopoulos and L.~Maiani, \emph{{Weak Interactions with
  Lepton-Hadron Symmetry}},
  \href{https://doi.org/10.1103/PhysRevD.2.1285}{\emph{Phys. Rev. D} {\bfseries
  2} (1970) 1285}.

\bibitem{Ciuchini:2015qxb}
M.~Ciuchini, M.~Fedele, E.~Franco, S.~Mishima, A.~Paul, L.~Silvestrini et~al.,
  \emph{{$B\to K^* \ell^+ \ell^-$ decays at large recoil in the Standard Model:
  a theoretical reappraisal}},
  \href{https://doi.org/10.1007/JHEP06(2016)116}{\emph{JHEP} {\bfseries 06}
  (2016) 116} [\href{https://arxiv.org/abs/1512.07157}{{\ttfamily
  1512.07157}}].

\bibitem{Belle-II:2023esi}
{\scshape Belle-II} collaboration, \emph{{Evidence for $B^+ \to K^+ \nu
  \bar{\nu}$ decays}},
  \href{https://doi.org/10.1103/PhysRevD.109.112006}{\emph{Phys. Rev. D}
  {\bfseries 109} (2024) 112006}
  [\href{https://arxiv.org/abs/2311.14647}{{\ttfamily 2311.14647}}].

\bibitem{Gubernari:2023puw}
N.~Gubernari, M.~Reboud, D.~van Dyk and J.~Virto, \emph{{Dispersive analysis of
  $B \to K^{(*)}$ and $B_{s} \to \phi$ form factors}},
  \href{https://doi.org/10.1007/JHEP12(2023)153}{\emph{JHEP} {\bfseries 12}
  (2023) 153} [\href{https://arxiv.org/abs/2305.06301}{{\ttfamily
  2305.06301}}].

\bibitem{Bouchard:2013eph}
{\scshape HPQCD} collaboration, \emph{{Rare decay $B \to K \ell^+ \ell^-$ form
  factors from lattice QCD}},
  \href{https://doi.org/10.1103/PhysRevD.88.054509}{\emph{Phys. Rev. D}
  {\bfseries 88} (2013) 054509}
  [\href{https://arxiv.org/abs/1306.2384}{{\ttfamily 1306.2384}}].

\bibitem{Bailey:2015dka}
J.A.~Bailey et~al., \emph{{$B\to K l^+l^-$ Decay Form Factors from Three-Flavor
  Lattice QCD}}, \href{https://doi.org/10.1103/PhysRevD.93.025026}{\emph{Phys.
  Rev. D} {\bfseries 93} (2016) 025026}
  [\href{https://arxiv.org/abs/1509.06235}{{\ttfamily 1509.06235}}].

\bibitem{Parrott:2022rgu}
{\scshape HPQCD} collaboration, \emph{{$B \to K$ and $D\to K$ form factors from
  fully relativistic lattice QCD}},
  \href{https://doi.org/10.1103/PhysRevD.107.014510}{\emph{Phys. Rev. D}
  {\bfseries 107} (2023) 014510}
  [\href{https://arxiv.org/abs/2207.12468}{{\ttfamily 2207.12468}}].

\bibitem{Kamenik:2009kc}
J.F.~Kamenik and C.~Smith, \emph{{Tree-level contributions to the rare decays
  $B^+ \to \pi^+ \nu \bar{\nu} $, $B^+ \to K^+ \nu \bar{\nu} $, and $B^+ \to
  K^{*+} \nu \bar{\nu} $ in the Standard Model}},
  \href{https://doi.org/10.1016/j.physletb.2009.09.041}{\emph{Phys. Lett. B}
  {\bfseries 680} (2009) 471}
  [\href{https://arxiv.org/abs/0908.1174}{{\ttfamily 0908.1174}}].

\bibitem{Parrott:2022zte}
{\scshape HPQCD} collaboration, \emph{{Standard Model predictions for $B\to K
  \ell^+ \ell^-$, $B \to K \ell^-_1 \ell^+_2$ and $B\to K \nu \bar{\nu}$ using
  form factors from $N_f = 2+1+1$ lattice QCD}},
  \href{https://doi.org/10.1103/PhysRevD.107.014511}{\emph{Phys. Rev. D}
  {\bfseries 107} (2023) 014511}
  [\href{https://arxiv.org/abs/2207.13371}{{\ttfamily 2207.13371}}].

\bibitem{Becirevic:2023aov}
D.~Be\v{c}irevi\'c, G.~Piazza and O.~Sumensari, \emph{{Revisiting $B\rightarrow
  K^{(*)} \nu {\bar{\nu }}$ decays in the Standard Model and beyond}},
  \href{https://doi.org/10.1140/epjc/s10052-023-11388-z}{\emph{Eur. Phys. J. C}
  {\bfseries 83} (2023) 252}
  [\href{https://arxiv.org/abs/2301.06990}{{\ttfamily 2301.06990}}].

\bibitem{Tian:2024ubt}
H.-J.~Tian, H.-B.~Fu, T.~Zhong, Y.-X.~Wang and X.-G.~Wu, \emph{{The rare decay
  $B^+ \to K^+\ell^+\ell^-(\nu\bar{\nu})$ under the QCD sum rules approach}},
  \href{https://arxiv.org/abs/2411.12141}{{\ttfamily 2411.12141}}.

\bibitem{Athron:2023hmz}
P.~Athron, R.~Martinez and C.~Sierra, \emph{{$B$ meson anomalies and large $B^+
  \to K^+ \nu \bar{\nu}$ in non-universal $U(1)'$ models}},
  \href{https://doi.org/10.1007/JHEP02(2024)121}{\emph{JHEP} {\bfseries 02}
  (2024) 121} [\href{https://arxiv.org/abs/2308.13426}{{\ttfamily
  2308.13426}}].

\bibitem{Bause:2023mfe}
R.~Bause, H.~Gisbert and G.~Hiller, \emph{{Implications of an enhanced $B\to K
  \nu \bar{\nu}$ branching ratio}},
  \href{https://doi.org/10.1103/PhysRevD.109.015006}{\emph{Phys. Rev. D}
  {\bfseries 109} (2024) 015006}
  [\href{https://arxiv.org/abs/2309.00075}{{\ttfamily 2309.00075}}].

\bibitem{Allwicher:2023xba}
L.~Allwicher, D.~Becirevic, G.~Piazza, S.~Rosauro-Alcaraz and O.~Sumensari,
  \emph{{Understanding the first measurement of $\mathcal{B}(B\to K \nu
  \bar{\nu})$}},
  \href{https://doi.org/10.1016/j.physletb.2023.138411}{\emph{Phys. Lett. B}
  {\bfseries 848} (2024) 138411}
  [\href{https://arxiv.org/abs/2309.02246}{{\ttfamily 2309.02246}}].

\bibitem{Abdughani:2023dlr}
M.~Abdughani and Y.~Reyimuaji, \emph{{Constraining light dark matter and
  mediator with $B^+\to K^+ \nu \bar{\nu}$ data}},
  \href{https://doi.org/10.1103/PhysRevD.110.055013}{\emph{Phys. Rev. D}
  {\bfseries 110} (2024) 055013}
  [\href{https://arxiv.org/abs/2309.03706}{{\ttfamily 2309.03706}}].

\bibitem{Chen:2023wpb}
C.-H.~Chen and C.-W.~Chiang, \emph{{Flavor anomalies in leptoquark model with
  gauged $U(1)_{L_\mu-L_\tau}$}},
  \href{https://doi.org/10.1103/PhysRevD.109.075004}{\emph{Phys. Rev. D}
  {\bfseries 109} (2024) 075004}
  [\href{https://arxiv.org/abs/2309.12904}{{\ttfamily 2309.12904}}].

\bibitem{He:2023bnk}
X.-G.~He, X.-D.~Ma and G.~Valencia, \emph{{Revisiting models that enhance $B^+
  \to K^+ \nu \bar{\nu}$ in light of the new Belle II measurement}},
  \href{https://doi.org/10.1103/PhysRevD.109.075019}{\emph{Phys. Rev. D}
  {\bfseries 109} (2024) 075019}
  [\href{https://arxiv.org/abs/2309.12741}{{\ttfamily 2309.12741}}].

\bibitem{Datta:2023iln}
A.~Datta, D.~Marfatia and L.~Mukherjee, \emph{{$B\to K\nu \bar{\nu}$, MiniBooNE
  and muon $g-2$ anomalies from a dark sector}},
  \href{https://doi.org/10.1103/PhysRevD.109.L031701}{\emph{Phys. Rev. D}
  {\bfseries 109} (2024) L031701}
  [\href{https://arxiv.org/abs/2310.15136}{{\ttfamily 2310.15136}}].

\bibitem{Altmannshofer:2023hkn}
W.~Altmannshofer, A.~Crivellin, H.~Haigh, G.~Inguglia and J.~Martin~Camalich,
  \emph{{Light new physics in $B \to K^{(*)} \nu \bar{\nu}$?}},
  \href{https://doi.org/10.1103/PhysRevD.109.075008}{\emph{Phys. Rev. D}
  {\bfseries 109} (2024) 075008}
  [\href{https://arxiv.org/abs/2311.14629}{{\ttfamily 2311.14629}}].

\bibitem{McKeen:2023uzo}
D.~McKeen, J.N.~Ng and D.~Tuckler, \emph{{Higgs portal interpretation of the
  Belle II $B^+ \to K^+ \nu \nu$ measurement}},
  \href{https://doi.org/10.1103/PhysRevD.109.075006}{\emph{Phys. Rev. D}
  {\bfseries 109} (2024) 075006}
  [\href{https://arxiv.org/abs/2312.00982}{{\ttfamily 2312.00982}}].

\bibitem{Fridell:2023ssf}
K.~Fridell, M.~Ghosh, T.~Okui and K.~Tobioka, \emph{{Decoding the $B \to K \nu
  \nu$ excess at Belle II: Kinematics, operators, and masses}},
  \href{https://doi.org/10.1103/PhysRevD.109.115006}{\emph{Phys. Rev. D}
  {\bfseries 109} (2024) 115006}
  [\href{https://arxiv.org/abs/2312.12507}{{\ttfamily 2312.12507}}].

\bibitem{Ho:2024cwk}
S.-Y.~Ho, J.~Kim and P.~Ko, \emph{{Recent $B^+ \to K^+ \nu \bar{\nu}$ Excess
  and Muon $g-2$ Illuminating Light Dark Sector with Higgs Portal}},
  \href{https://arxiv.org/abs/2401.10112}{{\ttfamily 2401.10112}}.

\bibitem{Chen:2024jlj}
F.-Z.~Chen, Q.~Wen and F.~Xu, \emph{{Correlating $B \to K^{(*)} \nu \bar{\nu}$
  and flavor anomalies in SMEFT}},
  \href{https://doi.org/10.1140/epjc/s10052-024-13425-x}{\emph{Eur. Phys. J. C}
  {\bfseries 84} (2024) 1012}
  [\href{https://arxiv.org/abs/2401.11552}{{\ttfamily 2401.11552}}].

\bibitem{Gabrielli:2024wys}
E.~Gabrielli, L.~Marzola, K.~M\"u\"ursepp and M.~Raidal, \emph{{Explaining the
  $B^+ \to K^+ \nu \bar{\nu}$ excess via a massless dark photon}},
  \href{https://doi.org/10.1140/epjc/s10052-024-12818-2}{\emph{Eur. Phys. J. C}
  {\bfseries 84} (2024) 460}
  [\href{https://arxiv.org/abs/2402.05901}{{\ttfamily 2402.05901}}].

\bibitem{Hou:2024vyw}
B.-F.~Hou, X.-Q.~Li, M.~Shen, Y.-D.~Yang and X.-B.~Yuan, \emph{{Deciphering the
  Belle II data on $ B\to K\nu \overline{\nu} $ decay in the (dark) SMEFT with
  minimal flavour violation}},
  \href{https://doi.org/10.1007/JHEP06(2024)172}{\emph{JHEP} {\bfseries 06}
  (2024) 172} [\href{https://arxiv.org/abs/2402.19208}{{\ttfamily
  2402.19208}}].

\bibitem{Chen:2024cll}
C.-H.~Chen and C.-W.~Chiang, \emph{{Rare B and K decays in a scotogenic
  model}}, \href{https://doi.org/10.1103/PhysRevD.110.075036}{\emph{Phys. Rev.
  D} {\bfseries 110} (2024) 075036}
  [\href{https://arxiv.org/abs/2403.02897}{{\ttfamily 2403.02897}}].

\bibitem{He:2024iju}
X.-G.~He, X.-D.~Ma, M.A.~Schmidt, G.~Valencia and R.R.~Volkas, \emph{{Scalar
  dark matter explanation of the excess in the Belle II $B^+ \to K^+ +$
  invisible measurement}},
  \href{https://doi.org/10.1007/JHEP07(2024)168}{\emph{JHEP} {\bfseries 07}
  (2024) 168} [\href{https://arxiv.org/abs/2403.12485}{{\ttfamily
  2403.12485}}].

\bibitem{Bolton:2024egx}
P.D.~Bolton, S.~Fajfer, J.F.~Kamenik and M.~Novoa-Brunet, \emph{{Signatures of
  light new particles in $B \to K^{(*)} E_\mathrm{miss}$}},
  \href{https://doi.org/10.1103/PhysRevD.110.055001}{\emph{Phys. Rev. D}
  {\bfseries 110} (2024) 055001}
  [\href{https://arxiv.org/abs/2403.13887}{{\ttfamily 2403.13887}}].

\bibitem{Marzocca:2024hua}
D.~Marzocca, M.~Nardecchia, A.~Stanzione and C.~Toni, \emph{{Implications of $B
  \rightarrow K \nu {\bar{\nu }}$ under rank-one flavor violation hypothesis}},
  \href{https://doi.org/10.1140/epjc/s10052-024-13534-7}{\emph{Eur. Phys. J. C}
  {\bfseries 84} (2024) 1217}
  [\href{https://arxiv.org/abs/2404.06533}{{\ttfamily 2404.06533}}].

\bibitem{Rosauro-Alcaraz:2024mvx}
S.~Rosauro-Alcaraz and L.P.S.~Leal, \emph{{Disentangling left and right-handed
  neutrino effects in $B \to K^{(*)} \nu \nu$}},
  \href{https://doi.org/10.1140/epjc/s10052-024-13104-x}{\emph{Eur. Phys. J. C}
  {\bfseries 84} (2024) 795}
  [\href{https://arxiv.org/abs/2404.17440}{{\ttfamily 2404.17440}}].

\bibitem{Eguren:2024oov}
J.F.~Eguren, S.~Klingel, E.~Stamou, M.~Tabet and R.~Ziegler, \emph{{Flavor
  phenomenology of light dark vectors}},
  \href{https://doi.org/10.1007/JHEP08(2024)111}{\emph{JHEP} {\bfseries 08}
  (2024) 111} [\href{https://arxiv.org/abs/2405.00108}{{\ttfamily
  2405.00108}}].

\bibitem{Buras:2024ewl}
A.J.~Buras, J.~Harz and M.A.~Mojahed, \emph{{Disentangling new physics in $K
  \to \pi \nu \bar{\nu}$ and $B \to K(K^*) \nu \bar{\nu}$ observables}},
  \href{https://doi.org/10.1007/JHEP10(2024)087}{\emph{JHEP} {\bfseries 10}
  (2024) 087} [\href{https://arxiv.org/abs/2405.06742}{{\ttfamily
  2405.06742}}].

\bibitem{Hati:2024ppg}
C.~Hati, J.~Leite, N.~Nath and J.W.F.~Valle, \emph{{QCD axion, color-mediated
  neutrino masses, and B+\textrightarrow{}K++Emiss anomaly}},
  \href{https://doi.org/10.1103/PhysRevD.111.015038}{\emph{Phys. Rev. D}
  {\bfseries 111} (2025) 015038}
  [\href{https://arxiv.org/abs/2408.00060}{{\ttfamily 2408.00060}}].

\bibitem{Allwicher:2024ncl}
L.~Allwicher, M.~Bordone, G.~Isidori, G.~Piazza and A.~Stanzione,
  \emph{{Probing third-generation New Physics with $K\to \pi \nu\bar{\nu}$ and
  $B\to K^{(*)} \nu\bar{\nu}$}},
  \href{https://doi.org/10.1016/j.physletb.2025.139295}{\emph{Phys. Lett. B}
  {\bfseries 861} (2025) 139295}
  [\href{https://arxiv.org/abs/2410.21444}{{\ttfamily 2410.21444}}].

\bibitem{Becirevic:2024iyi}
D.~Be\v{c}irevi\'c, S.~Fajfer, N.~Ko\v{s}nik and L.~Pavi\v{c}i\'c,
  \emph{{Right-handed interactions in puzzling $B$-decays}},
  \href{https://doi.org/10.1016/j.physletb.2025.139285}{\emph{Phys. Lett. B}
  {\bfseries 861} (2025) 139285}
  [\href{https://arxiv.org/abs/2410.23257}{{\ttfamily 2410.23257}}].

\bibitem{Altmannshofer:2024kxb}
W.~Altmannshofer and S.~Roy, \emph{{A joint explanation of the $B\to \pi K$
  puzzle and the $B \to K \nu \bar{\nu}$ excess}},
  \href{https://arxiv.org/abs/2411.06592}{{\ttfamily 2411.06592}}.

\bibitem{Buras:2024mnq}
A.J.~Buras and P.~Stangl, \emph{{On the Interplay of Constraints from $B_s$,
  $D$, and $K$ Meson Mixing in $Z^\prime$ Models with Implications for $b\to s
  \nu\bar\nu$ Transitions}},
  \href{https://arxiv.org/abs/2412.14254}{{\ttfamily 2412.14254}}.

\bibitem{Buras:2014fpa}
A.J.~Buras, J.~Girrbach-Noe, C.~Niehoff and D.M.~Straub, \emph{{$ B\to
  {K}^{\left(\ast \right)}\nu \overline{\nu} $ decays in the Standard Model and
  beyond}}, \href{https://doi.org/10.1007/JHEP02(2015)184}{\emph{JHEP}
  {\bfseries 02} (2015) 184} [\href{https://arxiv.org/abs/1409.4557}{{\ttfamily
  1409.4557}}].

\bibitem{Buchalla:1993bv}
G.~Buchalla and A.J.~Buras, \emph{{QCD corrections to rare K and B decays for
  arbitrary top quark mass}},
  \href{https://doi.org/10.1016/0550-3213(93)90405-E}{\emph{Nucl. Phys. B}
  {\bfseries 400} (1993) 225}.

\bibitem{Misiak:1999yg}
M.~Misiak and J.~Urban, \emph{{QCD corrections to FCNC decays mediated by Z
  penguins and W boxes}},
  \href{https://doi.org/10.1016/S0370-2693(99)00150-1}{\emph{Phys. Lett. B}
  {\bfseries 451} (1999) 161}
  [\href{https://arxiv.org/abs/hep-ph/9901278}{{\ttfamily hep-ph/9901278}}].

\bibitem{Buchalla:1998ba}
G.~Buchalla and A.J.~Buras, \emph{{The rare decays $K\to \pi \nu\bar\nu$, $B
  \to X \nu\bar\nu$ and $B \to l^+ l^-$: An Update}},
  \href{https://doi.org/10.1016/S0550-3213(99)00149-2}{\emph{Nucl. Phys. B}
  {\bfseries 548} (1999) 309}
  [\href{https://arxiv.org/abs/hep-ph/9901288}{{\ttfamily hep-ph/9901288}}].

\bibitem{Brod:2010hi}
J.~Brod, M.~Gorbahn and E.~Stamou, \emph{{Two-Loop Electroweak Corrections for
  the $K \to \pi \nu \bar{\nu}$ Decays}},
  \href{https://doi.org/10.1103/PhysRevD.83.034030}{\emph{Phys. Rev. D}
  {\bfseries 83} (2011) 034030}
  [\href{https://arxiv.org/abs/1009.0947}{{\ttfamily 1009.0947}}].

\bibitem{Detmold:2016pkz}
W.~Detmold and S.~Meinel, \emph{{$\Lambda_b \to \Lambda \ell^+ \ell^-$ form
  factors, differential branching fraction, and angular observables from
  lattice QCD with relativistic $b$ quarks}},
  \href{https://doi.org/10.1103/PhysRevD.93.074501}{\emph{Phys. Rev. D}
  {\bfseries 93} (2016) 074501}
  [\href{https://arxiv.org/abs/1602.01399}{{\ttfamily 1602.01399}}].

\bibitem{Altmannshofer:2017bsz}
W.~Altmannshofer, M.J.~Baker, S.~Gori, R.~Harnik, M.~Pospelov, E.~Stamou
  et~al., \emph{{Light resonances and the low-$q^{2}$ bin of $ {R}_{K^{*}} $}},
  \href{https://doi.org/10.1007/JHEP03(2018)188}{\emph{JHEP} {\bfseries 03}
  (2018) 188} [\href{https://arxiv.org/abs/1711.07494}{{\ttfamily
  1711.07494}}].

\bibitem{ParticleDataGroup:2024cfk}
{\scshape Particle Data Group} collaboration, \emph{{Review of particle
  physics}}, \href{https://doi.org/10.1103/PhysRevD.110.030001}{\emph{Phys.
  Rev. D} {\bfseries 110} (2024) 030001}.

\bibitem{Gubernari:2018wyi}
N.~Gubernari, A.~Kokulu and D.~van Dyk, \emph{{$B\to P$ and $B\to V$ Form
  Factors from $B$-Meson Light-Cone Sum Rules beyond Leading Twist}},
  \href{https://doi.org/10.1007/JHEP01(2019)150}{\emph{JHEP} {\bfseries 01}
  (2019) 150} [\href{https://arxiv.org/abs/1811.00983}{{\ttfamily
  1811.00983}}].

\bibitem{Horgan:2013hoa}
R.R.~Horgan, Z.~Liu, S.~Meinel and M.~Wingate, \emph{{Lattice QCD calculation
  of form factors describing the rare decays $B \to K^* \ell^+ \ell^-$ and $B_s
  \to \phi \ell^+ \ell^-$}},
  \href{https://doi.org/10.1103/PhysRevD.89.094501}{\emph{Phys. Rev. D}
  {\bfseries 89} (2014) 094501}
  [\href{https://arxiv.org/abs/1310.3722}{{\ttfamily 1310.3722}}].

\bibitem{Horgan:2015vla}
R.R.~Horgan, Z.~Liu, S.~Meinel and M.~Wingate, \emph{{Rare $B$ decays using
  lattice QCD form factors}},
  \href{https://doi.org/10.22323/1.214.0372}{\emph{PoS} {\bfseries LATTICE2014}
  (2015) 372} [\href{https://arxiv.org/abs/1501.00367}{{\ttfamily
  1501.00367}}].

\bibitem{Blake:2022vfl}
T.~Blake, S.~Meinel, M.~Rahimi and D.~van Dyk, \emph{{Dispersive bounds for
  local form factors in $\Lambda_b \to \Lambda$ transitions}},
  \href{https://doi.org/10.1103/PhysRevD.108.094509}{\emph{Phys. Rev. D}
  {\bfseries 108} (2023) 094509}
  [\href{https://arxiv.org/abs/2205.06041}{{\ttfamily 2205.06041}}].

\bibitem{BaBar:2013npw}
{\scshape BaBar} collaboration, \emph{{Search for $B \to K^{(*)} \nu \overline
  \nu$ and invisible quarkonium decays}},
  \href{https://doi.org/10.1103/PhysRevD.87.112005}{\emph{Phys. Rev. D}
  {\bfseries 87} (2013) 112005}
  [\href{https://arxiv.org/abs/1303.7465}{{\ttfamily 1303.7465}}].

\bibitem{Belle:2017oht}
{\scshape Belle} collaboration, \emph{{Search for $B\to h\nu\bar{\nu}$ decays
  with semileptonic tagging at Belle}},
  \href{https://doi.org/10.1103/PhysRevD.96.091101}{\emph{Phys. Rev. D}
  {\bfseries 96} (2017) 091101}
  [\href{https://arxiv.org/abs/1702.03224}{{\ttfamily 1702.03224}}].

\bibitem{Belle-II:2018jsg}
{\scshape Belle-II} collaboration, \emph{{The Belle II Physics Book}},
  \href{https://doi.org/10.1093/ptep/ptz106}{\emph{PTEP} {\bfseries 2019}
  (2019) 123C01} [\href{https://arxiv.org/abs/1808.10567}{{\ttfamily
  1808.10567}}].

\bibitem{Amhis:2023mpj}
Y.~Amhis, M.~Kenzie, M.~Reboud and A.R.~Wiederhold, \emph{{Prospects for
  searches of $ b\to s\nu \overline{\nu} $ decays at FCC-ee}},
  \href{https://doi.org/10.1007/JHEP01(2024)144}{\emph{JHEP} {\bfseries 01}
  (2024) 144} [\href{https://arxiv.org/abs/2309.11353}{{\ttfamily
  2309.11353}}].

\end{thebibliography}
\end{document}